\documentstyle[12pt,aasms4]{article}
{

\slugcomment{to appear in Astronomical Journal, 2001}

\lefthead{Rosado et al.}
\righthead{}

\begin{document}
\linespread{.8}
{
\title{ Kinematics of Herbig-Haro objects and jets in the Orion Nebula}

\author{M. Rosado\altaffilmark{1}, E. de la Fuente\altaffilmark{1,2}, L.
Arias\altaffilmark{1}, A. Raga\altaffilmark{1}, and E. Le
Coarer\altaffilmark{3}}

\altaffiltext{1}{Instituto de Astronom\'\i a,
UNAM, Apdo. Postal 70-264, 04510 M\'exico, D. F., M\'exico}
\altaffiltext{2}{Facultad de Ciencias,
UNAM,  M\'exico, D. F., C.P. 04510}
\altaffiltext{3}{Observatoire de Grenoble, BP53X Grenoble CEDEX, France}

\begin{abstract}
 
We have surveyed the inner 5\arcmin  ~ of the Orion Nebula by means of H$\alpha$ and [NII]
Fabry-Perot imaging spectroscopy to  present a kinematical study of the Herbig-Haro objects in the nebula. The objects studied in this work are HH 202, 203,
204, 529, 269 and other associated features. For HH 202
we find new features that, because of their high velocities (up to 100 km s$^{-1}$) indicate the presence of an outflow that probably is a HH flow not catalogued previously. HH 202 could be only a part of this larger outflow.
 Large internal motions are found in the fainter regions of  HH 203-204, as well as evidence of transverse density gradients that could account for the asymmetry in the brightness distribution of HH 204.  We report for
the first time a high blueshifted velocity ($-$118 km s$^{-1}$) associated with
HH 204, and show that  the apex of HH 204 is  indeed the zone of maximum velocity, 
in agreement with bow shock models. We also studied the radial velocity field of HH 269 finding features associated with the HH object. From our studies, we find kinematic evidence that suggests that HH 203-204 and HH 202 are part of a big ($\sim$ 0.55 pc)  bipolar HH outflow.  
\end{abstract}



\section{Introduction}

The Orion Nebula (M 42) was discovered in 1610 by Nicholas-Claude Fabri de Peiresc (Bigourdan 1916, Frommert \& Kronberg 2000). Later, in 1656, Christian Huygens rediscovered it and reported the inner region of this nebula describing, for the first time, the Trapezium stars (Huygens, 1659). This nebula is still being studied at several
wavelengths because Orion is one of the nearest star forming regions (Orion is located at a
distance of 450 pc) and consequently, the phenomena can be studied with the
highest spatial resolution. In the Orion region  several phases of
the interstellar medium coexist, such as molecular, neutral and ionized gas. With the development of near IR imaging and spectroscopy in recent years,  new
phenomena have been revealed that constitute important clues in the
understanding of the process of star formation and evolution of shocks in a
molecular medium (see for example, the H$_{2}$ fingers discovered near the
BN-IRc2 object in the OMC-1 molecular cloud whose bow shocks host several
Herbig-Haro objects, {\it {e.g.}} Axon \& Taylor 1984, Hartigan  {\it {et al}}.
1987 and Salas {\it {et al}}. 1999). The Orion Nebula is also
one of the more studied objects with the HST, which provides a unique set of
images in several emission lines revealing a plethora of new objects with the
best detail. A general information and historic description of this nebula are presented in de la Fuente (2001).

The kinematics of the Orion Nebula have been studied 
mainly with long-slit spectroscopy (Goudis {\it {et al}}. 1984,  Casta\~neda 1988, Wen \& O'Dell 1993). The long-slit data do not
give the spatial coverage required to study larger, possible high velocity
features as well as the whole set of interesting objects.  That is why  recent
Fabry-Perot spectroscopy at [SII] and [OIII] (O'Dell {\it {et al}}. 1997b) has been
successful in revealing high velocity flows  and proplyds.

Two other findings in the study of star forming regions 
motivated our interest in the Orion Nebula:

-- The discovery that Herbig-Haro (HH)  flows could be much larger than
previously thought (Reipurth {\it {et al}}. 1997) implying a revision of the energetics and
time scales involved in these flows. These giant HH flows are constituted of
two or more HH objects that, in the past, were thought to be isolated. 

-- The interest in studying jets produced by newly formed stars that are photo-ionized by
external sources. The most striking example is the recently discovered jet in
the Trifid Nebula (Cernicharo {\it {et al}}. 1997) whose peculiar kinematics has been
studied by Rosado {\it {et al}}. (1999). Theoretical models have been
recently developed to study the complex emission and kinematical properties of
these jets (Raga {\it {et al}}. 2000 ).

The Orion Nebula is the better place to study these phenomena. The large
quantity of HH objects discovered there and the fact that in this nebula coexist star
forming regions embedded in the UV flux of the Trapezium stars ensure that
these kinds of studies can be accomplished successfully with observations which
provide  large spatial coverage and high spectral resolution, qualities easily
furnished by imaging Fabry-Perot spectroscopy.

With the aim of obtaining a global view of the kinematics of the HH objects
and jets in the Orion Nebula, allowing us to look for large dimension structures  linking some of the objects, we have undertaken a study based on imaging Fabry-Perot spectroscopy in the emission lines of H$\alpha$, [NII], [SII] and [OIII]. We present results based on  H$\alpha$ and [NII]
observations. In Section 2 we comment on the observations and data reduction.
In Section 3 we give our results about Orion HH objects. In Section 4 we present morphological and kinematic evidences suggesting the existence of a big bipolar outflow that includes several of the studied HH objects. In Section 5  we give our conclusions.

\section{Observations and data reduction} 

The Fabry-Perot (FP) observations of the Orion Nebula
were carried out during several observing runs (though the more successful one was
during the nights from November 30 to December 5 1996)
at the f/7.5 Cassegrain focus of the 2.1~m telescope of the Observatorio
Astron\'omico Nacional at San Pedro M\'artir, B.C. (M\'exico) using the
UNAM Scanning Fabry-Perot Interferometer PUMA (Rosado {\it {et al}}. 1995).

A 1024$\times$1024 thinned Tektronix CCD detector, with
an image scale of 0.59\arcsec~pixel$^{-1}$, was used only in its central 512$\times$512 pixels to cover a
field of 5\arcmin~   in the sky.  Thus, the resulting image format was of
512$\times$512 pixels, with a spatial resolution of 0.59\arcsec~pixel $^{-1}$.

Interference filters centered at  H$\alpha$ and [NII]( $\lambda$  6583 \AA\ ),
 having  bandpasses of 20 and 10 \AA\ , respectively,  were used in order to
isolate the H$\alpha$ and [N~II] ($\lambda$~6583 \AA\ )  emission lines,
respectively. Other interference filters centered at [SII]( $\lambda$ 6720
\AA\ ) and [OIII]( $\lambda$~5007 \AA\ ), with  bandpasses of 20 and 10
\AA\ , respectively, were used to isolate the [S~II] and [O~III] emission
lines. As mentioned before, in this work we only present the study of the
H$\alpha$ and [NII] emission lines since O'Dell {\it {et al}}. (1997b) have
already published results of the  [SII] and [OIII] lines.

The scanning Fabry-Perot interferometer is an ET-50 of Queensgate
Instruments  with a servo-stabilization system. The main characteristics
of this interferometer are: interference order of 330, free spectral range of
19.89 \AA\ (equivalent to a velocity range of 908
km~s$^{-1}$) and sampling spectral resolution of 0.41 \AA\
(equivalent to 18.9 km~s$^{-1}$), at H$\alpha$, achieved by 
scanning the interferometer gap at 48 positions. Thus, the resulting data
cubes have dimensions of 512$\times$512$\times$48. The velocity resolution
of the data is two times the sampling resolution, i. e., the velocity resolution is of 37.8 km~s$^{-1}$. On the other hand, the accuracy in
the measurement of the peak velocities of the velocity profiles is of about 5 km~s$^{-1}$.

With these setups, we have obtained two nebular data cubes at H$\alpha$
and six  at [NII] with
exposure times of 24 min. each.  The nebular cubes of the same emission line
were co-added in order to
enhance the S/N ratio of the faint regions, resulting in two nebular data
cubes with  total  exposure times of 48 and 144 min.,  respectively. We also
obtained  calibration data cubes spaced at the beginning, during  and at the
end of the observations in order to check for possible flexures of the
equipment. For the calibration cubes we have used the H$\alpha$ line (from a H
lamp) so that we do not have to do any phase shift correction due to the FP
interferometer plates for the H$\alpha$ data cubes while we needed to do such
a correction for the [NII] data cubes.

The data reduction and analysis were performed using the specific reduction
package CIGALE (Le Coarer {\it {et al}}. 1993). This software was used to remove
 cosmic rays, to carry out the wavelength calibration
of the data cubes, to obtain continuum subtracted velocity-cubes 
and to carry out the emission line profile analysis. We  have also
used some {IRAF\altaffilmark{2}} routines for part of the data reduction.

Since the Orion Nebula is so bright, some problems are present when observing it,
as discussed in O'Dell {\it {et al}}. (1997b). We took two or more 
exposures instead of a single longer one in order to detect the bright Trapezium
stars without saturation. We have identified quite well the reflective ghosts
due to those bright stars. An inspection of them during the course of the
observations allowed us to place the orientation of the field in the less harmful way with the ghosts falling outside the more interesting regions. On the
other hand, while we also observed the ``ringing'' of the calibrated velocity
maps discussed in O'Dell {\it {et al}}. (1997b) --due to the undersampling
of the FP scans because of the increase of accuracy in the velocities given the high S/N of the observations of  bright emission lines-- we do not
obtain appreciable velocity changes or spurious features in the radial velocity
profiles in order to do the interpolations that O'Dell {\it {et al}}. (1997b)
have developed to handle their FP data. This is because
our reduction packages work with the original interference rings and  not with
the calibrated velocity maps in the extraction of radial velocity profiles. In
fact, calibrated velocity maps are used only to study the morphology of the
emission at different radial velocities. For these velocity maps some ringing
is present.

\altaffiltext{2}{ IRAF is distributed by the National Optical Astronomy Observato
ries, operated by the Association of Universities for Research in Astronomy, Inc.
, under cooperative agreement with the National Science Foundation.}

\section{ Kinematics of individual HH objects}

Figures 1 and 2 show the velocity maps at $V_{helio}$ = -- 127 km
s$^{-1}$ (i.e., ~ 145 km s$^{-1}$ blueshifted relative to the velocity of the intense nebular
background) at H$\alpha$ and [NII], respectively. In Figure 1 we have
identified the different  detected objects: the HH objects HH 202, HH 269
(barely seen at H$\alpha$), HH 203 and HH 204, the E-W jet mentioned in O'Dell {\it {et al}}. (1997b) and several
proplyds which we have denominated following the terminology of
O'Dell {\it {et al}}. (1993) and O'Dell \& Wen (1994).

\subsection{ HH 202}

HH 202 was discovered by Cant\'o {\it {et al}}. (1980) as an emission line object showing two knots embedded in an arc-shaped nebulosity. These authors reported that the northern knot moves at $V_{helio}$ = -- 31 km s$^{-1}$.  Further spectroscopic work, in [OIII],
of  the HH 202 region (Meaburn 1986) showed that the arc-shaped nebulosity
was not seen in [OIII] and that a more extensive region (a half-disk of 20\arcsec~
diameter), emitting in the [OIII] lines, had similar velocities to those of
the  knot seen in [SII].  

Further spectroscopy has been
performed in several emission lines by O'Dell {\it {et al}}. (1991)  showing two velocity systems  that these authors identified with a
bow shock and its corresponding Mach disk. More recently, O'Dell {\it {et
al}}. (1997a) have obtained wonderful HST  images of the HH 202 region in the
[SII], H$\alpha$ and [OIII] lines. O'Dell {\it {et al}}.
(1997a) also detect strong [OIII] emission inside the arc-shaped nebulosity.
 O'Dell
{\it {et al}}. (1997b) obtained [SII] and [OIII] FP spectroscopy of the 
Trapezium region (and consequently, of the HH 202 region). They found a
blueshifted portion of HH 202 that extends towards the NW. In the case of
HH 202 the identification of the driving source is still a mystery given the richness of the Orion Nebula field, which has a large amount of possible candidates.

No reliable proper motions have been measured for this object. Indeed, Bally
{\it {et al}}. (2000) report proper motion determinations of several objects in the Orion Nebula but HH 202 is not included. Their proper motion measurements were based on the comparison of 1994 and 1998/99 HST monochromatic images of Orion.  Previous proper motion determinations
were done based on continuum images where the emission lines were not isolated.
Since in this work we show that the structure and kinematics of HH 202 are very dependent on the studied line, it is possible that those determinations are spurious. At least, they are not confirmed by the recent study of Bally
{\it {et al}}. (2000).

Figure 3 shows close-ups  of the H$\alpha$ and [NII] velocity maps at 
$V_{helio}$ = -- 90 km s$^{-1}$  showing the region around HH 202 (which is
located at the NW corner of the FP field). The [NII] image shows the arc-shaped
nebulosity ending to the SW  in one of the HH 202 knots. On the other hand, the H$\alpha$ image shows several morphological
differences both in the HH object and in the neighboring region. We detect the
arc-shaped nebulosity but the southern knot is not easily disentangled
from the emission of a bright  nebulosity and of several other knots (as confirmed by an
[OIII] image not shown). Instead, we see   a bright head of irregular shape
(somewhat like an arrow head) pointing in the E-W  direction,  with three faint
filamentary extensions also oriented in the E-W direction. The longest
filamentary extensions seem to form a rotated spur, or $\Omega$.

We also find that the
arc-shaped nebulosity of HH 202 seems to be part of a larger scale lobe (of 82\arcsec $\times$ 25\arcsec~ and aperture angle of $\sim$ 40\arcdeg),  oriented in the NW-SE direction, that converges in
its SE end towards a point close to the eastern region of the E-W jet of
O'Dell {\it {et al}}. (1997a). Furthermore, inside this lobe, two elongated
cavities (one of 35\arcsec $\times$ 9\arcsec~  and another of 55\arcsec $\times$ 12\arcsec, with aperture angles of $\sim$ 20\arcdeg) somewhat similar to bow
shocks or to the H$_{2}$ Orion's fingers (Axon \& Taylor 1984, Salas {\it {et
al}}. 1999), are detected.  On the other
hand, the walls of the lobe are better appreciated in [NII], but the
elongated interior cavities (hereafter called `fingers') are confused with the bright background
nebula.

We extracted radial velocity profiles integrated over boxes
of  different dimensions selected in order to cover the different features mentioned above. The
velocity profiles are complex and cannot be fitted by a simple Gaussian
function. Figure 4 shows a typical velocity profile integrated over the box shown in Figure 3. In general, the
brightest component corresponds to the intense HII region and has a gradual
velocity variation with position from $V_{helio}$ = +10 km s$^{-1}$ (for the zone near to
the E-W jet) to $V_{helio}$ = +17 km s$^{-1}$ (for the region of the HH 202 head). In
addition, the velocity profiles show wings blueshifted relative to the bright
HII region component, reaching relative velocities of up to 100 km s$^{-1}$. Blueshifted
wings are found: at the E-W jet ($V_{relative}$ = 70 km s$^{-1}$), in the `fingers' 
($V_{relative}$ = 60 km s$^{-1}$), in the cavity or lobe ($V_{relative}$ = 86 km s$^{-1}$), in the
arc  ($V_{relative}$ = 78 km s$^{-1}$), in the northern knot  ($V_{relative}$ = 70 km s$^{-1}$), in
the southern knot  ($V_{relative}$ = 70 km s$^{-1}$) and in the spur  ($V_{relative}$ = 105
km s$^{-1}$). From the profile decomposition, we were not able to find any redshifted
component as suggested in the velocity maps. Thus, we can conservatively
say that motions of up to  100 km s$^{-1}$ are present in this system and that the spur
(which has the largest velocity values) belongs to the HH system. The high velocities found in these features indicate the presence of an HH flow of large dimensions ( $\sim$ 0.18 pc long) not catalogued as
such because of the difficulties of disentangling it from the bright nebular
background. It is unclear whether this flow is associated with
HH 202 or whether it constitutes another HH system.

\subsection{ HH 203 and HH 204}

These objects were discovered by Munch \& Wilson (1962).  Taylor \&
Munch (1978) show images of these
bow-shaped objects near the star $\Theta^{2}$ A Ori, and also report the
existence of a third object, `object C', located in the opposite side to
$\Theta^{2}$ A  Ori (this object has recently been identified as the proplyd 244--440). These authors also give radial velocities and velocity
dispersions of the several different features in these objects, obtained with a
Coud\'e spectrograph and a tandem scanning FP interferometer. However, Taylor
\& Munch (1978) interpreted the velocity motions of the different features as
due to an expanding motion of a wind blown bubble.  Cant\'o {\it {et al}}. (1980) identified these objects as HH objects. Walsh (1982) obtained the electron
densities of HH 203 and 204. O'Dell {\it {et al}}. (1997a) show HST images of these HH objects.

Proper motions for HH 204 have been obtained by
Cudworth \& Stone (1977),  using broad bandpass images, indicating tangential velocities between 30 and 70
km s$^{-1}$ directed towards the apex of the bow. Hu (1996) has also
measured tangential velocities of 0 and 70 km s$^{-1}$ for HH 203 and HH 204,
respectively, using HST  images in the [NII] line before and after the refurbishment mission. The more recent (and more accurate) work of Bally {\it {et al}}. (2000),
using HST monochromatic images obtained after the refurbishment mission, gives as a result tangential velocities of 74 $\pm$ 10 km s$^{-1}$ for HH 203 and of 59 $\pm$ 6 km s$^{-1}$ for HH 204, moving toward a position angle of 140\arcdeg.

There are at least two important questions related to these objects: are HH 203 and HH 204 parts of the same object? and, why does HH 204 show an asymmetry in its brightness distribution?.

 In Figures 5 and  6 we show some
representative velocity maps, at H$\alpha$ and [NII], respectively, where the most important features appear. There appears to be some differences in the morphology of HH 203 and HH 204 according to the emission line (H$\alpha$ or [NII]) and the velocity.

With respect to HH 204, 
we see that, in H$\alpha$, HH 204 has a complete and symmetrical conical or bow shock shape (height of at least 34\arcsec~ and aperture angle of  $\sim$
 65\arcdeg) whose
apex correspond to a bright knot (B according to Taylor \& Munch 1978). The cone's  H$\alpha$ emission starts to be confused with the bright nebular background at the Orion bar, while in [NII]
the cone is still detected well above the `bar' (to the NW). We distinguish a pronounced asymmetry in the brightness distribution of the cone: the
side near to  the star $\Theta^{2}$ A Ori is brighter than the side located  away from this star. This asymmetry is better appreciated in [NII] than in  H$\alpha$. Also, the
whole conical shape of HH 204 is better appreciated at extreme velocities
(either redshifted or blueshifted: $V_{helio}$ = +43 km s$^{-1}$ and -- 50 km s$^{-1}$).

 HH 203, featuring a jet-like appearance of at least 83\arcsec~
   long, appears to be partially superposed on HH 204. It is
better disentangled from HH 204's cone at blueshifted velocities ($V_{helio}$ =
   -- 70
km s$^{-1}$) and its H$\alpha$ emission is found in both sides of the `bar'. The jet-like feature encompassing HH 203 is directed towards HH 202 and the E-W jet
discussed in the previous section. The end of this jet-like object (knot A1 according to
Taylor \& Munch 1978) is bent and brighter.  In [NII], we distinguish well the jet-like feature but, 
 it seems to be shorter than in H$\alpha$  (no emission is found in the western side of the bar) and it appears to have an interaction with the neighboring cone wall of HH 204 just to the north of
the `bar'. Thus, HH 203 seems to be a jet colliding with the bow shock of HH 204. There are well known examples of jets entering bow shocks on the side, as  the cases of HH 34 (Buehrke {\it {et al}}. 1988) and HH 47 (Eisloeffel \& Mundt 1994). A precessing jet could be responsible of this particular morphology.

Figure 7 shows a close-up of the H$\alpha$ velocity map at $V_{helio}$ = -- 89
km s$^{-1}$ containing HH 203 and HH 204. Superimposed on this map are the boxes over which we extracted radial velocity profiles (both
from the H$\alpha$ and [NII] cubes).
The radial velocity profiles integrated over the boxes are, in general, complex and cannot
be fitted by a single Gaussian function. Indeed, they can be fitted by: (1) a
narrow and bright component (that corresponds to the HII region emission, with
velocity widths varying between 20 and 40 km s$^{-1}$) with $V_{helio}$
varying between +13 to +25 km s$^{-1}$ (the highest values are found
in preference near the `bar' and to the North of it) and (2) blueshifted wings which are  fainter and broader than the redshifted component  (typical velocity
width of 70 km s$^{-1}$). In the brightest zones (the end of HH 203 and
the apex of HH 204) we detect a splitting of the velocity profiles into two components
of more or less the same intensity. In those cases, the redshifted components are clearly distinguished from the HII region and, consequently, this component should belong to the HH objects.

In Figure 7 we also show the velocity difference  of the components that fit the radial velocity profiles integrated over the boxes, or in other words, the object velocities relative to the HII region. The brightest zone of HH 203 shows a splitting of the velocity profiles in two
components  at +13 and -- 53 km s$^{-1}$ while the apex of HH 204 shows another
splitting in two components at +20 and -- 24 km s$^{-1}$ and, in addition, a wing
reaching $V_{helio}$ = -- 118 km s$^{-1}$. The splitting values found for these zones are in
agreement with several other determinations (Taylor \& Munch 1978, Cant\'o
{\it {et al}}. 1980, Walsh 1982 and Bally {\it {et al}}. 2000). However, the high blueshifted velocity
value for the wing in the velocity profile of the apex HH 204 has never been
reported before. It is important because the apex of HH 204 is thus, the zone
with maximum velocity, as predicted by bow shock models.  

On the other hand, the zones with maximum velocities relative to the HII region
correspond  to the  cone side of HH 204 far away from $\Theta^{2}$ A Ori (relative
blueshifted velocities of up to 100 km s$^{-1}$) while in the bright side of the cone, near to
$\Theta^{2}$ A Ori, the relative velocities are of 70 km s$^{-1}$. Consequently, the
fainter regions have larger velocities. Henney   (1996) has proposed that a transverse density gradient in the ambient medium where a bow shock propagates, could lead to an asymmetry in brightness of the bow shock. Our results give some support to this idea  of a transverse density gradient because we find that there is a slight velocity gradient running perpendicular to the axis of HH 204 in the sense that the
fainter (and less dense) regions  have larger velocities.

  Finally, HH 203 and HH 204 seem to be part of a structure of large dimensions or lobe. Indeed, a careful inspection of Figures 5 and 6 suggests the detection of an incomplete lobe ending at HH 204 on one side, and at the mark shown in Figures 5 and 6  on the other side.  This lobe would be 132\arcsec~  or 0.29 pc long and is also oriented in the NW-SE direction as the lobe associated with HH 202. It is  more intense in its northern half. The possible existence of this lobe is also revealed in  O'Dell's image of the Orion Nebula published in the National Geographic Supplement (Grosvenor {\it {et al}}. 1995).

\subsection{ HH 269 and other Herbig-Haro objects near the Trapezium stars}

\subsubsection{ HH 269 }

HH 269 has an elliptical shape of 41\arcsec $\times$ 32\arcsec~  with the major axis
oriented in the E-W direction. A distinctive feature of this object is that two knots are detected (hereafter called the E and W knots, respectively) at the ends of the major axis.

The singular appearance of HH 269 and the possibility that it could be an HH 
object were reported for the first time by Feibelman (1976), who also
 estimated its tangential expansion velocity  (90 - 100 km s$^{-1}$). However, given that the proper motion estimates were made from plates obtained from different telescopes, these results should be regarded with caution. Later
on, Walter {\it {et al}}. (1995) reported their results of HST imagery and low
and high resolution spectroscopy showing that this object is of low
excitation, obtaining the electron densities of the knots and the central
region and detecting blueshifted velocity components at $V_{helio}$ = -- 22.5 and -- 13
km s$^{-1}$ for the W and E knots, respectively. These authors suggest that the
central region has an emission line spectrum more of the type of an HII region
than of an HH object. The more recent work of Bally {\it {et al}}. (2000)
shows HST imagery of this object revealing that the E and W knots are filamentary and embedded in the tenuous elliptical nebula. These authors measure tangential velocities of 56  km s$^{-1}$ at P.A. =  250\arcdeg for the E knot and 71 km s$^{-1}$ at  P.A. =  260\arcdeg for the W knot. 

We have analized our [NII] velocity cubes in order to know more about this
object. Figure 8 shows the [NII] velocity maps where HH 269 is detected.
As we can see, the object becomes brighter in the maps corresponding to $V_{helio}$
= -- 14 and +5 km s$^{-1}$ and, consequently, its systemic velocity must be between
that velocity range and thus, blueshifted relative to the systemic velocity of
the HII region. Also, the object shows more knots that the previously reported: the E and W knots
already mentioned and, at least, five additional knots. Figure 9 is a close-up of the velocity map at $V_{helio}$ = +5 km s$^{-1}$. Superimposed on it we show several
boxes where we have extracted integrated  FP velocity profiles. Also, in this figure
 the heliocentric radial velocities found from our profile fitting are displayed. 

We find a bright and narrow component (FWHM = 20  km s$^{-1}$) present all over the whole extension of the object, 
which we interpret as due to the HII region emission. Its velocities
vary from $V_{helio}$ = -- 1 to +6 km s$^{-1}$. For the W knot we find a splitting in the
velocity profile into two components: the HII region component and another
component at $V_{helio}$ = -- 33 km s$^{-1}$ (slightly larger than the value reported by
Walter {\it {et al}}. 1995). However,
we do not find the blueshifted component at -- 13 km s$^{-1}$ reported by Walter {\it {et al}}.
(1995) for the E knot. This negative result is probably  due to our lower spectral
resolution that does not allow us to disentangle this component from the
bright HII region component.

On the other hand, the central and southern zones show the HII region velocity
component and very faint blueshifted wings of up to -- 190 km s$^{-1}$.  The wings are visualized in the velocity maps as arcs running perpendicular to the E-W axis of HH 269. Given the high blueshifted velocities
we think that these arcs are related to the HH object.  

\subsubsection{ Other HH objects }

We have identified in the FP field of view of our observations the regions where the HH objects: HH 528, HH 529, HH 530 and HH 507, reported in  Bally {\it {et al}}. (2000), are located in order to cross-identify these objects and obtain their radial velocities. In the case of HH 528, no conspicuous feature could be seen neither in the H$\alpha$ nor in the [NII]  maps at high velocities. In the case of HH 530, although we detect it in the [NII] FP cubes, the integrated radial velocity profiles of several knots and regions of this nebulosity only show the velocity component corresponding to the HII region. HH 529 is different
because it is quite conspicuous at H$\alpha$ where it can be seen, at blueshifted velocities, to the East of the E-W jet (Figure 3). In our ground-based velocity maps, the bow shock shown in the HST image of  Bally {\it {et al}}. (2000) appears as a bright elongated knot with faint curved extensions. The radial velocities of the E-W jet have been discussed in Section 3.1 and, in  Bally {\it {et al}}. (2000), there is no measured proper motion for this interesting feature. The radial velocity profiles we have obtained for the tip of the bow shock of HH 529 and its southern wall show that, in addition to the HII region component at  $V_{helio}$ = +15 km s$^{-1}$, there is a blueshifted component at  $V_{helio}$ = -- 19 km s$^{-1}$ at the tip of the bow shock. On the other hand, Bally {\it {et al}}. (2000) report proper motions of several features in HH 529 implying transverse velocities of 85 $\pm$ 24  km s$^{-1}$ with a P.A. = 100\arcdeg, i.e., directed eastwards from the E-W jet.

\section{ The large scale features}

Our results show that HH 202 seems to be part of a lobe which is larger than the known dimensions of the arc-shaped nebulosity associated with HH 202. This lobe is blueshifted relative to the main HII region velocity and shows the large internal motions (of up to 100 km s$^{-1}$) characteristic of HH flows.  The [NII] velocity cubes show that this lobe (hereafter the NW lobe) is formed by two `fingers' starting from a region close to the E-W jet discovered by O'Dell  {\it {et al}}. (1997a). The point of convergence of this large flow is marked in Figure 3.

On the other hand, HH 203 and HH 204 also seem to be part of a large structure or lobe (hereafter the SE lobe) of similar dimensions to the ones of the NW lobe. HH 204 is at one of the ends of the SE lobe while the other end is located close to the E-W jet, as in the case of the NW lobe. Furthermore, the SE lobe shares the same orientation and point of convergence of the flow as the NW lobe
(see Figure 5).

The other HH objects we have detected and the additional ones revealed by their high transverse velocities ( Bally {\it {et al}}. 2000) are not so large or not so symmetrical relative to the point of convergence of the large structures forming HH 202 and HH 203-204.  Bally {\it {et al}}. (2000) have suggested that at least six relatively large (~0.2 pc size) HH outflows are emerging from the OMC1-S cloud core and the sources CS-3 and FIR-4. However,considering the point of convergence that we have found for the ends of the NW and SE lobes, we see that the smallest HH objects are not symmetrical relative to this point. Indeed, in Figure 3 we have marked the end of the NW lobe (point of convergence of the flow) as well as several sources in the vicinity of OMC1-S (the IR sources A, B and C and the exciting sources of molecular outflows CS-3 and FIR-4 discussed in  Bally {\it {et al}}. 2000). As we can see from this figure, those sources are about 40 $\arcsec$ to the west of the center of convergence of the NW lobe. Furthermore, the proper motion vectors of HH 269 are oriented pointing towards a point to the north of OMC1-S for it is probable  that HH 269 does not belong to the same flow as the system HH 202 and HH 203-204.

Considering the results presented in the previous sections and in a preliminary study of our data (Rosado {\it {et al}}. 2001), we suggest that HH 202 and HH 203-204, are part of a large bipolar outflow, 0.55 pc long, that arises from an object close to the E-W jet. In addition, we have seen that the NW lobe is blueshifted relative to the background nebula while the SE lobe is also detected at redshifted velocities, as is typical for bipolar outflows. The measured proper motions of HH 203-204  ( Bally {\it {et al}}. 2000) agree with this interpretation while there are no reliable proper motions measurements for HH 202.
On the other hand, the region close to the E-W jet is so rich in objects that it is difficult to identify, by means of the existing stellar data, the object that could be the source of this suggested bipolar outflow. It is interesting to note that at the position of the suggested source there is a point-like source with high [NII]/H$\alpha$ line-ratio (see Figure 3). Figure 10 shows the H$\alpha$ velocity map at $V_{helio}$ = -- 50 km s$^{-1}$ of the inner 5\arcmin
~  of the Orion Nebula, which allows us to have a global view of this outflow. At blueshifted velocities, the NW lobe is better detected than the SE lobe. However, jet-like features (such as HH 203 and the `fingers' inside the NW lobe) are detected inside these lobes. Further studies of the stellar content close to the E-W jet and of proper motions of the HH 202 knots are required in order to confirm or reject this possibility.

\section{Conclusions}

We present a kinematic study of the Herbig-Haro objects HH
202, 203, 204, 269 and 529 using  H$\alpha$ and [NII] Fabry-Perot imaging spectroscopy.

For HH 202
we find new features that could belong to this HH object or that perhaps are associated with a different outflow. Because of its high velocity (of up to 100 km s$^{-1}$) this outflow can probably be an HH flow not catalogued previously.

We have found that HH 203 has a jet-like appearance at blueshifted velocities while HH 204 has a conical shape resembling a bow shock with a strong asymmetry in its brightness distribution. Large internal motions are found in the fainter regions of  HH 203-204, as well as some evidence of transverse density gradients. We show that the apex of HH 204 is the zone of maximum velocity in agreement with bow shock models.

We also studied the radial velocity field of HH 269 finding that near the center and to the south, a high blueshifted velocity component which is seen as a series of two arcs or bows running perpendicular to the main axis of this HH object. 

We searched for violent motions in other HH objects detected in the field of view and we have only found them for HH 529.

Finally, from the studies of the individual HH objects and of their close environments, we find kinematic evidence to suggest that HH 202 and HH 203-204  are part of a big bipolar ($\sim$ 0.55 pc)  HH outflow.

\section*{Acknowledgements}

The authors wish to acknowledge the comments of the anonymous referee and also, they thank  Alfredo D\'iaz and Carmelo Guzm\'an
for the computer help.  
MR wishes to acknowledge  the financial support from grants 400354-5-2398PE of CONACYT and  IN104696 of DGAPA-UNAM and EdelaF akcnowledges the finantial support from CONACYT grant 124449  and DGEP-UNAM through graduate scholarships.

\clearpage

\vfill\eject

\figcaption{  H$\alpha$ velocity map at  $V_{helio}$ = -- 127 km
s$^{-1}$  of the Orion Nebula obtained with the PUMA FP observations. Some of the important stars, HH objects and proplyds are marked.  
\label{fig1}}

\figcaption{ [NII]  velocity map at  $V_{helio}$ = -- 127 km
s$^{-1}$  of the Orion Nebula obtained with the PUMA FP observations.
\label{fig2}}

\figcaption{Close-ups  of the velocity maps at 
$V_{helio}$ = -- 90 km s$^{-1}$  showing the field near HH 202 which is
located at the NW corner. The arrow to the SE corresponds to one of the ends of the lobe. a) at  H$\alpha$ and b) at [NII]($\lambda$ 6583 \AA). The rectangular box marked in a) shows the region over which the radial velocity profiles, presented in Figure 4, were integrated. The star marks shown in a) represent the positions of the 2 $\micron$ sources (A, B and C) and of the molecular outflow sources CS-3 and
FIR-4 suggested in Bally  {\it {et al}}. (2000) as possible exciting sources.
\label{fig3}}

\figcaption{ Typical  a) H$\alpha$ and b) [NII] radial velocity profiles integrated over a zone of HH 202. The zone of integration is shown in Figure 3.
\label{fig4}}

\figcaption{ Close-ups  of representative H$\alpha$ velocity maps  
 showing the field near HH 203 and HH 204 which are located near the SE corner. The arrow to the NW corresponds to one of the ends of the lobe and it is at the same position as in Fig. 3. The heliocentric radial velocities ( in km
s$^{-1}$ ) are shown at the lower left corner. The box in the map at $V_{helio}$ = -- 51 km s$^{-1}$ corresponds to the field displayed in Figure 7.
\label{fig5}}

\figcaption{ The same as in Fig. 5 but for [NII]($\lambda$ 6583 \AA). 
\label{fig6}}

\figcaption{ Close-up  of the  H$\alpha$ velocity map at 
$V_{helio}$ = -- 89 km s$^{-1}$ showing HH 203 and HH 204. Radial velocity profiles were extracted by integrating over the boxes shown superimposed on this image. The numbers inside the boxes correspond to the difference in velocities (in  km s$^{-1}$) between the HII region component and  the blueshifted component for each integrated radial velocity profile.
\label{fig7}}

\figcaption{ Close-ups  of representative [NII]($\lambda$ 6583 \AA) velocity maps  
 showing the field of HH 269. The heliocentric radial velocities ( in km
s$^{-1}$ ) are shown at the lower right corner.
\label{fig8}}

\figcaption{ Close-up of the velocity map at $V_{helio}$ = +5 km s$^{-1}$ of HH 269. Several
boxes where  the FP velocity profiles were integrated are shown superimposed. The heliocentric radial velocities found from our profile fitting are also displayed. The high blueshifted velocities correspond to the wings in the velocity profiles mentioned in the text. 
\label{fig9}}

\figcaption{  H$\alpha$ velocity map, at $V_{helio}$ = -- 50 km s$^{-1}$, of the inner 5\arcmin ~ of the Orion Nebula. At this blueshifted velocity, the global view of the NW lobe, the E-W jet and the jet-like appearance of HH 203 are appreciated. The small nebulosity to the W of the E-W jet (marked with an arrow in Figures 3 and 6) corresponds to the location of the possible exciting source of the bipolar flow.
\label{fig10}}

}

\end{document}